\documentclass{article}
\usepackage[margin=2cm]{geometry}
\usepackage{amssymb}
\usepackage{natbib}
\begin{document}

\title{Open problem:\\Computing the optimal deterministic
  budget-balanced mechanism\thanks{Originally prepared for the program
    ``Economics and Computation'' held in the Fall 2015 semester at
    the Simons Institute for the Theory of Computing.}}
\author{Vincent Conitzer\\ \texttt{conitzer@cs.cmu.edu} } \date{}
\maketitle

\section{The problem}

{\bf Given:}
\begin{itemize}
\item a (constant-size) set of agents $\{1, \ldots, k\}$ (say, $k=2$)
\item for each agent $i$, a finite (explicitly enumerated) set of types $\Theta_i$ with a prior
  distribution $P_i$ (say, uniform)
\item a finite (explicitly enumerated) set of outcomes (e.g., allocations) $O$
\item for each agent $i$, a utility function $u_i: \Theta_i \times O
  \rightarrow \mathbb{Q}$
\end{itemize}

A {\em deterministic (direct-revelation) mechanism} consists of
a function $f: \Theta_1 \times \ldots \times \Theta_k \rightarrow O$
selecting an outcome, and, for each agent $i$, a function $\pi_i: \Theta_1
\times \ldots \times \Theta_k \rightarrow \mathbb{Q}$ selecting a payment
to that agent (negative if the agent pays).\\

\noindent
{\bf Problem:} Can an optimal (in terms of maximizing expected social
welfare) {\bf deterministic} ex-post individually rational,
dominant-strategy incentive compatible (i.e., strategy-proof), strongly
budget balanced mechanism be found in polynomial time?  That is, can the
following mixed integer linear program be solved in polynomial time?  (Let
$P$ denote the joint prior, $\theta$ a vector of types, and $x(\theta,o)$
the probability of choosing outcome $o$ for type vector $\theta$, necessarily in
$\{0,1\}$ in this context.  Note the payments $\pi_i(\theta)$ are also
variables, which do not need to take integer values.)\\

\noindent
maximize $\sum_\theta P(\theta) \sum_i \sum_o u_i(\theta_i,o)x(\theta,o)$\\
subject to\\
for all $\theta$: $\sum_i \pi_i(\theta) = 0$ (strong budget balance)\\
for all $i$ and $\theta$: $\pi_i(\theta) + \sum_o
u_i(\theta_i,o)x(\theta,o) \geq 0$ (ex-post IR)\\ 
for all $i$, $\theta$, and $\theta_i'$: $\pi_i(\theta) + \sum_o
u_i(\theta_i,o)x(\theta,o) \geq  \pi_i(\theta_{-i}, \theta_i') + \sum_o
u_i(\theta_i,o)x((\theta_{-i},\theta_i'),o)$ (strategy-proofness)\\
for all $\theta$: $\sum_o x(\theta,o)=1$\\
for all $\theta, o$: $x(\theta,o) \in \{0,1\}$\\

\noindent {\bf Variants of the problem:} could also consider (combinations
of): ex-interim IR, Bayes-Nash incentive compatibility, weak budget balance
(the mechanism is allowed to run a surplus {\bf but} this comes at a cost
to the objective, i.e., the money is burned rather than given to an outside
party).

\section{Further background}

We introduced the algorithmic study of such explicitly represented finite
mechanism design problems in~\citet{Conitzer02:Mechanism}, and studied it
in several other papers (e.g.,~\citet{Conitzer04:Self}).  As we point out
in these papers, if the mechanism is allowed to {\bf randomize}, then these
types of problem can all be solved in polynomial time (for a constant
number of agents) because they correspond to a linear program (simply drop
the integrality constraints above). Costis' second talk in the bootcamp
involved the same linear programming approach for a discretized mechanism
design problem, but he focused on scaling to larger numbers of agents
rather than on deterministic mechanisms.  

Still, in many real applications, randomized mechanisms would not be
acceptable.  Unfortunately, computing an optimal {\bf deterministic}
mechanism is NP-hard in many cases.  For example, finding the determinstic
mechanism that optimizes social welfare without using any payments
whatsoever is NP-hard, even if only a single agent reports a type (which
implies that it does not matter whether ex-interim or ex-post IR or IC is
used).  On the other hand, if we can use payments {\bf and} there is no
strong budget balance constraint (and payments do not count as a loss to
welfare, e.g., they're received by a seller) then the problem is trivially
polynomial-time solvable using VCG.  Chapter 6 of my
dissertation~\citep{Conitzer06:PhD} has reductions for various related
problems.

Incidentally, the variant where we allow the mechanism to run a surplus but
consider this surplus a loss from the perspective of welfare leads to some
highly nontrivial mechanism design problems
(e.g.,~\citet{Guo07:Worst,Moulin09:Almost,Guo08:Better,Clippel09:Destroy}).

\end{document}